\documentclass[a4paper,12pt]{article}
\usepackage{mathrsfs}
\usepackage{graphicx} 
\usepackage{epstopdf}
\usepackage{subfigure}
\usepackage{color}
\usepackage{ulem}

\begin{document}

\hoffset = -1truecm \voffset = -2truecm \baselineskip = 10 mm
\title{\bf Revealing mysteries in gamma-ray bursts: the role of gluon condensation}

\author{Wei Zhu$^a$\footnote{Corresponding author, E-mail: wzhu@phy.ecnu.edu.cn},  Xu-Rong Chen$^b$ and
Yu-Chen Tang$^c$  \\\\
         \normalsize $^a$Department of Physics, East China Normal University, Shanghai 200241, P.R. China\\
         \normalsize $^b$Institute of Modern Physics, Chinese Academy of Sciences, Lanzhou 730000,P.R. China\\
         \normalsize $^c$Key Laboratory of Dark Matter and Space Astronomy, Purple Mountain Observatory,\\
         \normalsize Chinese Academy of Sciences, Nanjing 210008, P.R. China}
\date{}

\newpage

\maketitle

\vskip 3truecm

\begin{abstract}

    We use a newly recognized gluon distribution in the nucleon, which was predicted by a QCD evolution equation to
consistently explain several intriguing phenomena associated with gamma-ray bursts. They are the GeV-TeV spectra of GRB 221009A, the remarkably symmetrical explosion cloud in kilonova AT2017gfo, and the absence of a very high-energy gamma-ray signature in GRB 170817A. We find that these occurrences can be attributed to the gluon condensation within nucleons, i.e., a significant number of soft gluons within nucleons are condensed at a critical momentum, resulting in the emergence of a steep and high peak in the gluon distributions. Through this profound connection between microscopic and macroscopic phenomena, we have not only expanded the applications of the hadronic scenario in cosmic gamma-ray emissions but also presented new evidence for the existence of gluon condensation.

\end{abstract}

{\bf keywords}: Gamma-ray bursts; GRB 221009A; AT2017gfo/GRB 170817A/GW170817; Gluon condensation

\newpage

\vskip 1truecm

\section{INTRODUCTION}

     Gamma-ray bursts (GRBs) represent the most potent and luminous electromagnetic events observed in distant galaxies
since the early universe. Long-duration GRBs are theorized to originate from the collapse of massive stars, while short-duration bursts are believed to result from the merger of two compact celestial objects, such as neutron stars or black holes. Despite the numerous theoretical models proposed to explain both long and short GRBs, the fundamental question surrounding these phenomena remains an enigma. To exemplify, we present two recent instances as follows.

    Recent reports of TeV spectra have opened a completely new energy window, expected to advance studies in this field.
For instance, a highly energetic outburst initially detected by Swift-XRT [1] was promptly confirmed as GRB 221009A by Fermi-GBM [2,3] and Fermi-LAT [4,5]. Unfortunately, space-based detectors, limited in their resolution capabilities to gamma-ray signals up to the 10 GeV energy level, faced constraints. However, the Large High Altitude Air Shower Observatory (LHAASO) detectors overcame this limitation, recording very high-energy (VHE) photons in the TeV energy range from GRB 221009A within the next 30 minutes [6]. The extensive array of ground-based detectors enabled an unprecedented level of precision, revealing that the gamma-ray brightness of GRB 221009A exceeded that of any previously observed GRB by 50 times. This extraordinary event has provided us with a more comprehensive VHE spectrum of the brightest GRB ever detected. It is conceivable that this long-duration GRB originated from the gravitational collapse of a massive star, leading to the formation of either a neutron star or a black hole. Upon the creation of a neutron star or black hole, the resultant formation of powerful particle jets containing protons, nuclei, and electrons occurs. These particles undergo acceleration, reaching velocities approaching the speed of light, primarily driven by shock waves. Subsequently, they release high-energy and VHE photons into space while traversing through remnants of the progenitor stars. The spectra of GRB 221009A present a distinctive structure in the GeV-TeV band, which is a new clue in unraveling the mystery of GRBs. Despite the numerous theoretical models proposed to elucidate these high-energy events, the true nature of the VHE radiation source remains elusive.

    An illustrative example of a short-duration GRB is the kilonova AT2017gfo [7], originating from the gravitational wave
event GW 170817 [8], also referred to as short-duration GRB 170817A due to the observation of electromagnetic (EM) counterparts [9]. This event marked the first multi-messenger observation related to the coalescence of two neutron stars. Upon scrutinizing the spectra of AT2017gfo, researchers made a surprising discovery: during its early stages, this kilonova exhibited a perfectly spherical shape [10]. This contradicts the expected characteristics of ultra-dense matter, as hydrodynamical models suggested that the resulting explosion cloud should have a flattened shape. This finding implies that the merger of two compact neutron cores occurred with minimal resistance. The specific mechanism underlying this spherical explosion remains a mystery.

    Gamma-rays in cosmic rays generally stem from two sources [11]: (i) the leptonic scenario, involving curvature radiation
or inverse Compton scattering of high-energy electrons, accompanied by low-energy (KeV) synchrotron radiation spectra, and (ii) the hadronic scenario, occurring in a two-step process: proton collisions initially produce a large number of secondary particles, with the pion meson being chief among them, and the neutral $\pi^0$ decays electromagnetically, resulting in a pair of photons. It is reasonable to consider that the hadronic scenario should dominate as a mechanism. This is because protons and nuclei carry most of the kinetic energy in the primary particle jets and are also accelerated by shock waves. Additionally, their radiation loss is minimal. However, in GRB studies, the application of the hadronic scenario is restricted. The reason is as follows.

  A proton consists of quarks and gluons. In high-energy proton collisions, hundreds or even thousands of secondary
particles, predominantly $\pi$ meson, can be generated. As the collision energy increases, an increasing number of gluons with small values of $x$ within the proton (where $x$ denotes the ratio of the longitudinal momentum of the gluon to that of the proton) contribute to the production of secondary particles. Consequently, gluons play a dominant role in pion production. The abundance of gluons rises steadily as $x$ decreases, without forming a distinct peak in the meson energy spectrum. This finding aligns with experimental observations at the Large Hadron Collider (LHC), currently the largest high-energy collider. On the other hand, the electromagnetic decay of the neutral $\pi^0$ results in a peak in the gamma-ray spectrum
at $m_{\pi}/2$, which is shifted to nearly 1 GeV due to the increased production of pions. Consequently, the energy spectra of the conventional hadronic scenario consistently cluster around $\sim 1~GeV$ in $E_{\gamma}^2\Phi_{\gamma}$ (referred to as the $\pi$-bump) [12]. Therefore, the hadronic scenario appears to be less appealing for independently explaining TeV emission in GRBs. The case of GRB 221009A has prompted us to revisit the structure of the proton. It is conceivable that the energy of the proton may surpass the acceleration capabilities of the LHC, potentially leading to a new understanding of the nature of gluons within the proton.
   .
    In this paper, our focus is on the hadronic scenario, with consideration given to a novel understanding of nucleon
structure. According to the evolution equations in quantum chromodynamics (QCD), gluon distributions are predicted to gradually approach an equilibrium state characterized by the splitting and fusion of gluons, a phenomenon known as the color glass condensate (CGC) [13], although this does not represent a true physical condensation. A further development of the evolution equations is the Zhu-Shen-Ruan (ZSR) equation [14,15], which suggests that at high energy, the evolution of gluon distributions in nucleons exhibits chaotic behavior, resulting in extremely strong shadowing and antishadowing effects. This leads to the accumulation of numerous gluons within a narrow phase space defined by a critical momentum $(x_c, k_c)$. This phenomenon is referred to as gluon condensation (GC) (see Fig. 1) [16,17].

    \begin{figure}[htbp]
    	\centering
        \subfigure{

    		\includegraphics[width=0.5\textwidth]{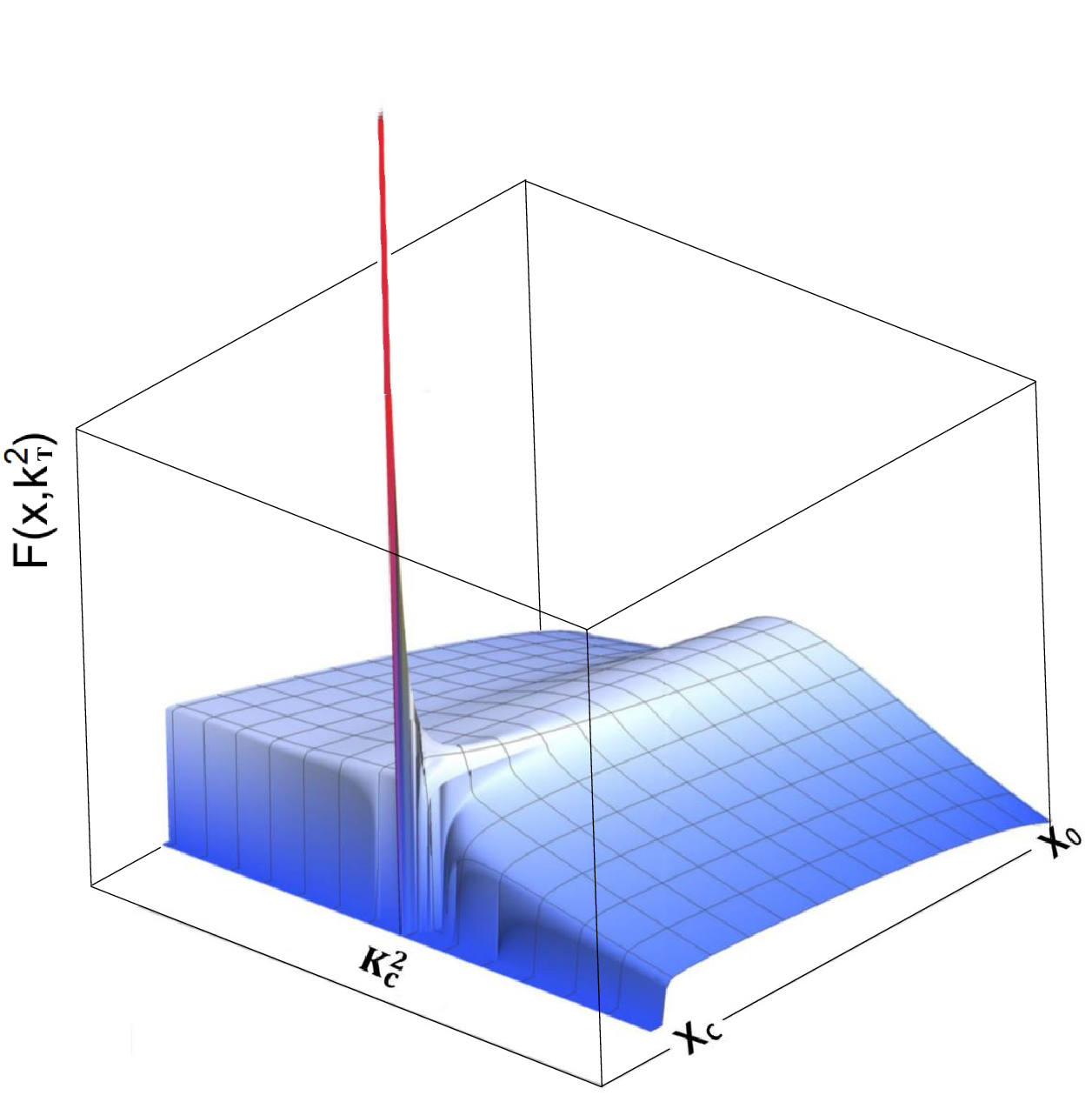}}
    	   	\caption{The evolution of gluon distribution in a QCD evolution equation [16,17] from the CGC
    		to GC, where gluons at $x<x_c$ are condensed at a critical momentum $(x_c,k_c)$. All coordinates are on the
            logarithmic scale. There are two characteristics of this distribution: a sharp peak at the critical momentum and no gluons present at $x<x_c$. We will show their different performances in GRB 221009A and GRB 170817A, respectively.
    	}\label{fig:1}

    \end{figure}

    A substantial distortion in the gluon distribution function, as suggested by the GC-effect, would likely have been noticeable
in proton colliders. The absence of its detection can be attributed to the limitation that the energy levels at the LHC are insufficient for the gluon peaks at $(x_c,k_c)$ to have reached the interaction zone. This realization prompted us to explore cosmic rays, where protons can be accelerated to energy levels surpassing those achievable at the LHC. The approach is straightforward and can be computed by directly substituting the distribution illustrated in Fig. 1 into the conventional hadronic framework. A concise derivation can be found in Sec. 2, with a more detailed explanation provided in [18-20].

    In Section 3, we demonstrate that the resulting gamma-ray spectrum conforms to a broken power-law (BPL) distribution,
featuring an attenuation factor in the tail. We refer to the traditional hadronic scenario incorporating the gluon condensation effect as the GC-model, its spectrum exhibits exceptional agreement with the GRB 221009A spectrum.

    The concept of GC not only involves the accumulation of gluons in the vicinity of $(x_c, k_c)$ but also implies
the disappearance of gluons smaller than $x_c$. Consequently, when the proton-proton (pp) collision energy surpasses the threshold $s_{cut}$, its total scattering cross-section is significantly suppressed, rendering the colliding protons transparent. Given that the threshold $s_{cut}$ is very high for pp and pA collisions, observing this phenomenon becomes challenging. An exception may be binary neutron star mergers, as the neutron star nucleus, being infinitely heavy, can have a sufficiently low $S_{cut}$ making the merger drag-free. We will qualitatively discuss the kilonova AT2017gfo event in Sec. 4.
Finally, the discussions a summary are given in Sec. 5.

\section{THE GC-MODEL}

    The GC-model is a combination of the traditional hadronic scenario and the GC-effect in the nucleon. One can be expected that
when the condensation peak of the gluon distribution in the proton shown in Fig. 1 enters the $pp$ interaction region, the resulting pion multiplicity will increase sharply, even reaching a saturation state. Surprisingly, we can prove that it leads to the BPL of the gamma spectrum using only energy conservation and relativistic covariance, which distinguishes the GC-model from other models.
For the convenience of readers' understanding this model, we systematically reconstruct the derivations in [18-20].

    In the hadronic scenario, approximately half of
the energy from parent protons is absorbed by valence quarks during a $pp$ collision, which subsequently forms the leading particles. The residual energy undergoes conversion into secondary hadrons, predominantly pions, within the central region. Subsequently, $\pi_0\rightarrow 2\gamma$ takes place. The spectrum of these photons in the Lab. system is computed by [11]

$$\Phi_{\gamma}(E_{\gamma})=C_{\gamma}\left(\frac{E_{\gamma}}{\mathrm{GeV}}\right)^{-\beta_{\gamma}}
\int_{E_{\pi}^{min}}^{\infty}dE_{\pi}
\left(\frac{E_p}{\mathrm{GeV}}\right)^{-\beta_p}N_{\pi}(E_p,E_{\pi})
\frac{d\omega_{\pi-\gamma}(E_{\pi},E_{\gamma})}{dE_{\gamma}},
\eqno(2.1)$$where the spectral index $\beta_\gamma$ incorporates the energy loss due to photon absorption by the medium, and
$N_\pi$ is the multiplicity of pions, $C_\gamma$ is a normalization constant that includes the kinematic
factor and the flux dimension. As customary, the accelerated protons are assumed to follow a simple
power law (PL) $N_p \propto E_p^{-\beta_p}$ near the source.

    For calculating $N_{\pi}$, we substitute the GC-distribution in Fig. 1 to the multiplicity $N_g$ of gluon mini-jet [21,22]

$$\frac{dN_g}{dk_T^2dy}=\frac{64N_c}{(N^2_c-1)k_T^2}\int q_T d
q_T\int_0^{2\pi}
d\phi\alpha_s(\Omega)\frac{F(x_1,\frac{1}{4}(k_T+q_T)^2)F(x_2,\frac{1}{4}
(k_T-q_T)^2)}{(k_T+q_T)^2(k_T-q_T)^2}, \eqno(2.2)$$where
$\Omega=Max\{k_T^2,(k_T+q_T)^2/4, (k_T-q_T)^2/4\}$; The longitudinal
momentum fractions of interacting gluons are fixed by kinematics
$x_{1,2}=k_Te^{\pm y}/\sqrt{S}$.

\begin{figure}[htbp]
	\begin{center}
		\includegraphics[width=0.8\textwidth]{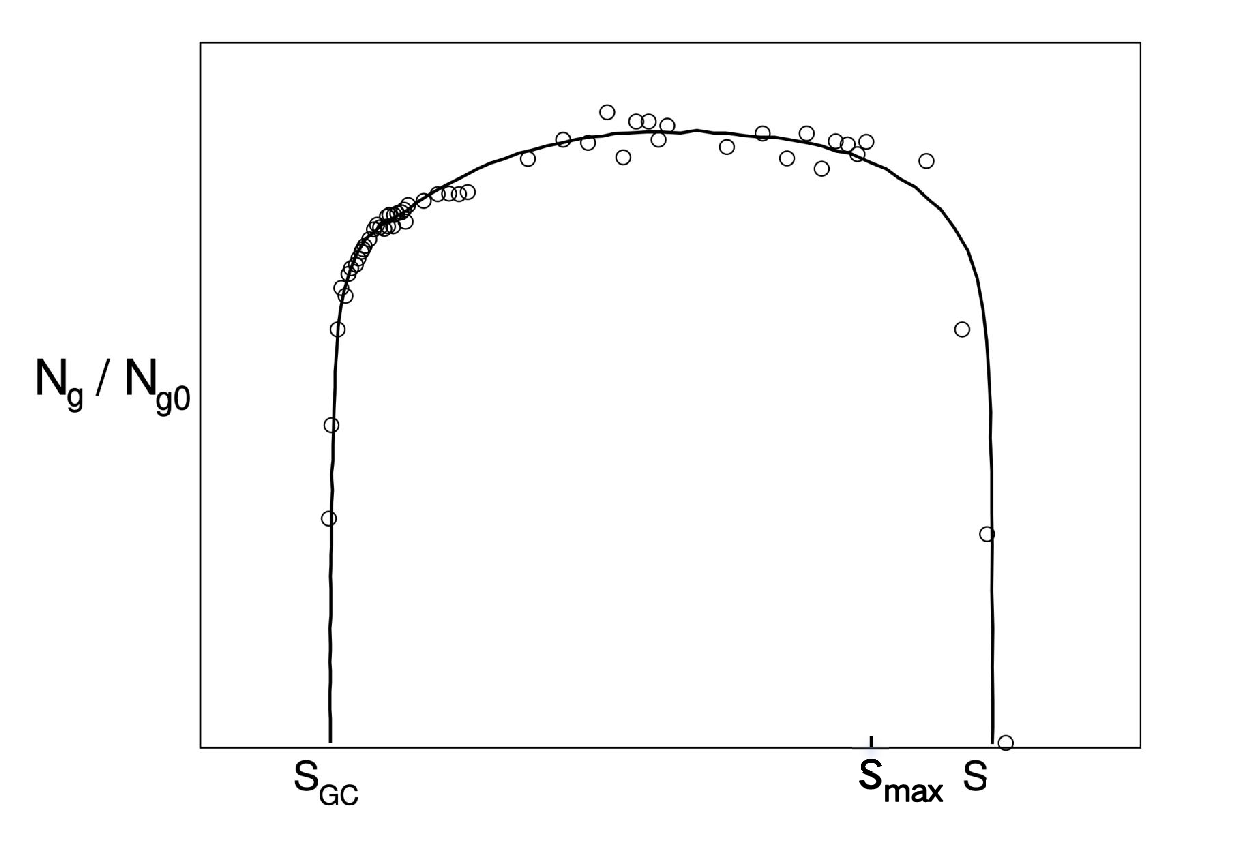} 
		\caption{ A schematic diagram for the ratio of the multiplicity of gluon mini-jets with and without the GC-effect. The empty circles are the solutions of Eq. (2.2) using the input in Fig. 1 and the solid curve is their smoothed result. All scales are taken as logarithmic unit.
		}\label{fig:2}
	\end{center}
\end{figure}

    As anticipated, an evident surge in $N_g$ becomes apparent when $\sqrt{S}$ surpasses $\sqrt{S_{GC}}$, signifying
the active involvement of the gluon distribution peaks in the generation of secondary particles (see Fig. 2). Nevertheless, this augmentation ceases beyond a certain higher energy threshold, leading to a rapid decline in meson yield to zero. This decline occurs due to the unavailability of gluons smaller than $x_c$. The resulting high plateau effect in the meson yield distinctly surpasses the modest peak associated with $\pi^0$ decay, contributing to a bulge in the gamma-ray spectrum.

    In the subsequent stage, we encounter $N_{\pi}$. The relationships among $N_{\pi}$, $E_p$, and $E_{\pi}$ in Eq. (2.1)
are highly intricate, since the energy-momentum of $\pi$ is randomly distributed. Conventionally, a multi-parameter empirical formulation is employed to articulate $N_{\pi}$. However, when addressing cosmic rays, whose energies may surpass the scale of the LHC, these empirical formulas are non-existent.

    Fortunately, QCD evolution equations can be used to predict these unknown knowledge, in particular,
the GC-effect can be employed to simplify this relationship. Experimental observations indicate that
as collision energy increases, $N_{\pi}$ also increases due to the greater participation of gluons in the process of generating lot of new particles. Recent research by Roberts et al. suggests that the effective hadron mass might be dynamically generated as a result of gluon nonlinearities [23]. We envisage that when a substantial number of condensed gluons at the threshold $x_c$ suddenly participate in the $pp$-collisions, it inevitably leads to a dramatic increase in the production of secondary meson. Since meson has mass, their yield $N_{\pi}$ is inherently constrained. In accordance with quantum theory, newly generated particles through collisions are initially excited from off-mass-shell to on-mass-shell, acquiring relative kinetic energy from the remaining interaction energy. In principle, the condensed gluons engaging in collisions may simultaneously generate a considerable number of secondary off-mass-shell meson at a given interaction energy. If these mesons are excited to on-mass-shell, they are capable of saturating all available energy, indicating that nearly all kinetic energies in collisions at the center-of-mass (C.M.) frame are utilized in creating the rest pions. This results in almost no relative momentum for the newly-formed meson, leading to the maximum value of $N_{\pi}$. While the validity of this saturation approximation will be scrutinized by subsequent observed data, adopting this limit allows us to circumvent the complex hadronization mechanism.

\begin{figure}[htbp]
	\begin{center}
		\includegraphics[width=0.8\textwidth]{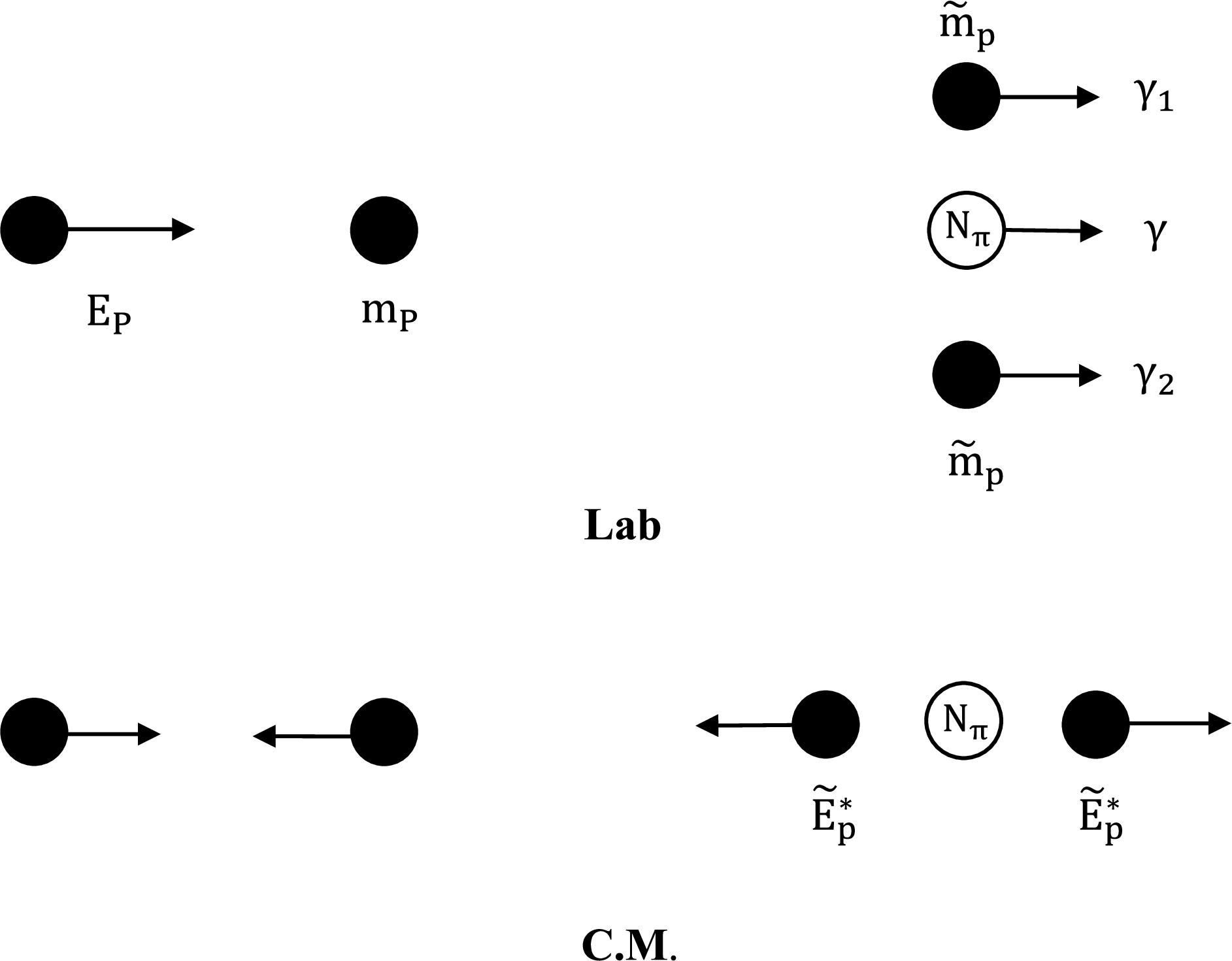} 
		\caption{$pp\rightarrow N_{\pi}$ in the different systems, where almost all available kinetic energies of collisions at the
			C.M. system are used to create pions due to the GC-effect.
		}\label{fig:3}
	\end{center}
\end{figure}

      We simplify the process of proton collisions producing secondary hadrons $pp\rightarrow X$ as $pp\rightarrow pp+N_{\pi}$,
where pions dominate the secondary hadrons and they are in a stationary state at the C.M. system due to the GC-effect.

      Energy conservation in Fig. 3 is written

$$E_p+m_p~~[in~the~Lab~system]=\tilde{m}_p\gamma_1+\tilde{m}_p\gamma_2+N_{\pi}m_{\pi}\gamma~~[in~the~C.M. system],  \eqno(2.3)$$where $\tilde{m}_p$ marks the leading particle and
$\gamma_i$ is the Lorentz factor. On the other hand,
the square of relativistic invariant total energy $S=(p_1+p_2)^2$ in the two systems is

$$S=2m_p^2+2E_pm_p~~[in~the~Lab~system]=(2\tilde{E}^*_p+N_{\pi}m_{\pi})^2~~[in~the~C.M. system]. \eqno(2.4)$$

    The proportion of the total collision energy occupied by the leading particles and the central particles is independent
of the selection of the reference frame. Using a following empirical relation [24], we remove the physical quantities about the leading particles in Eqs. (2.3) and (2.4)

$$2\tilde{E}^*_p=\left(\frac{1}{k}-1\right)N_{\pi}m_{\pi}, \eqno(2.5)$$and

$$\tilde{m}_p\gamma_1+\tilde{m}_p\gamma_2=\left(\frac{1}{k}-1\right)N_{\pi}m_{\pi}\gamma, \eqno(2.6)$$ $k\simeq 1/2$ is the inelasticity.
Thus, we have

$$\sqrt{2m_p^2+2E_pm_p}=\frac{1}{k}N_{\pi}m_{\pi}, \eqno(2.7)$$and

$$E_p+m_p=\frac{1}{k}N_{\pi}m_{\pi}\gamma, \eqno(2.8)$$

    We appoint that the energy in both equations above is measured in GeV units. After taking the
logarithm on both sides of the equations, using Eq. (2.7) and $\tilde{E}_p\equiv E_p+m_p$, we have

$$\frac{1}{2}\ln(2\tilde{E}_pm_p)=\ln\left(\frac{1}{k}N_{\pi}m_{\pi}\right), \eqno(2.9)$$or

$$\ln N_{\pi}=0.5\ln (\tilde{E}_p/GeV)+a, \eqno(2.10)$$where $a\equiv 0.5\ln (2m_p/GeV)-\ln (m_{\pi}/GeV)+\ln k$.

    Similarly, Eq. (2.8) leads to

$$\ln \tilde{E}_p=\ln\left(\frac{1}{k}N_{\pi}m_{\pi}\gamma\right)=\ln N_{\pi}+\ln m_{\pi}+\ln \gamma-\ln k=\frac{1}{2}
\ln \tilde{E}_p+\frac{1}{2}\ln (2m_p)+\ln \gamma,  \eqno(2.11)$$where Eq. (2.10) is used. Therefore,

$$\ln \gamma=\frac{1}{2}\ln \tilde{E}_p-\frac{1}{2}\ln (2m_p). \eqno(2.12)$$

    Since $E_{\pi}=m_{\pi}\gamma$, we have

    $$\ln E_{\pi}=\ln\gamma+\ln m_{\pi}=\frac{1}{2}\ln \tilde{E}_p-\frac{1}{2}\ln (2m_p)+\ln m_{\pi}=\frac{1}{2}\ln \tilde{E}_p+b',
    \eqno(2.13)$$with $b'=\ln m_{\pi}-1/2\ln (2m_p).$ Using Eq. (2.10) we have

    $$\ln N_{\pi}=\ln (E_{\pi}/GeV)+a-b'=\ln (E_{\pi}/GeV)+b, \eqno(2.14)$$where $b\equiv \ln (2m_p/GeV)-2\ln (m_{\pi}/GeV)+\ln k$.

     Thus, under the saturation condition for $N_{\pi}$, only two universal conditions (relativistic covariance and conservation
of energy) can determine the pion yield, which is a straight line in double logarithmic coordinates, i.e., the typical PL in $E_{\gamma}\in [E_{\pi}^{GC}, E_{\pi}^{max}]$ in Fig. 4.

\begin{figure}[htbp]
	\begin{center}
		\includegraphics[width=0.8\textwidth]{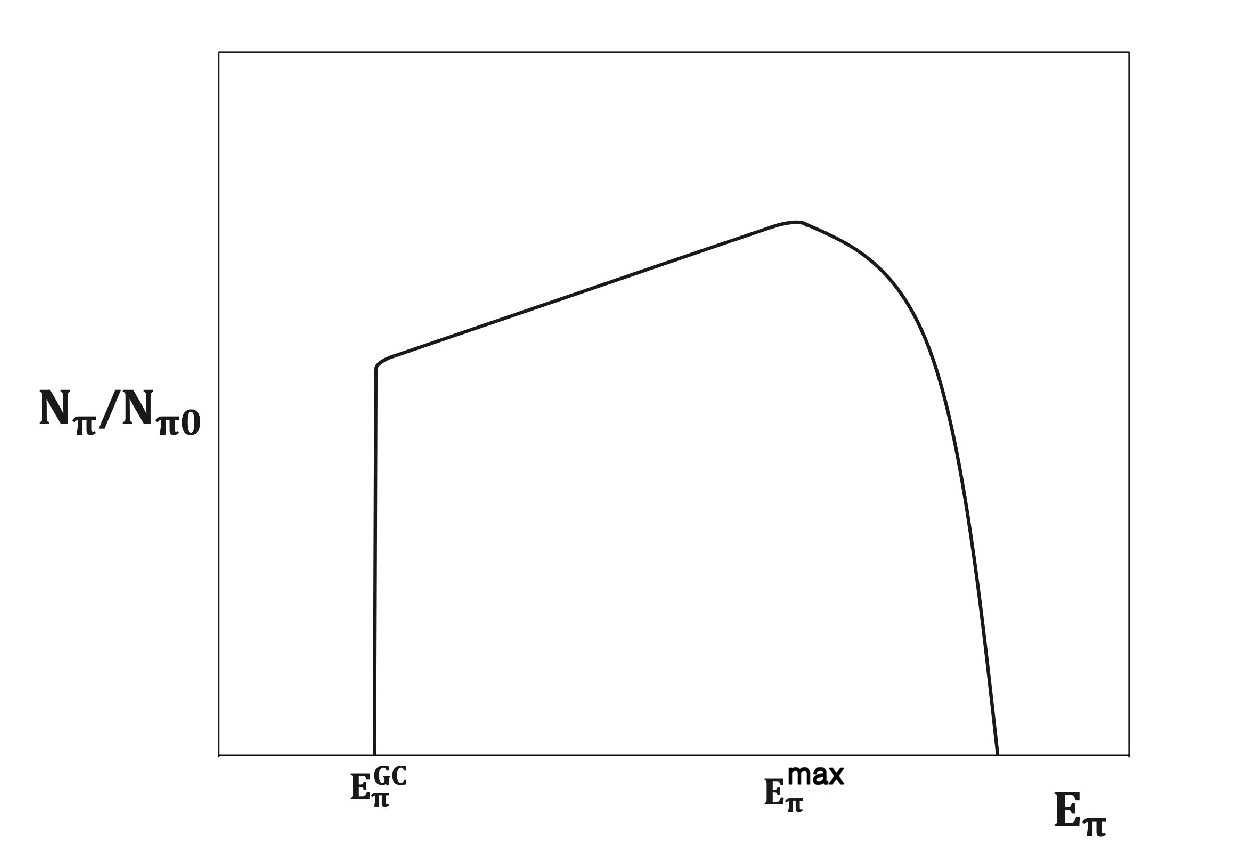} 
		\caption{ A schematic diagram for the ratio of the multiplicity of pion with and without the GC-effect. Note that the
			fluctuations in Fig. 2 have disappeared due to Eqs. (2.10) and (2.14).  All scales are taken as logarithmic units.
		}\label{fig:4}
	\end{center}
\end{figure}

    Substituting a standard spectrum of $\pi^0\rightarrow 2\gamma$ into Eq. (2.1), i.e.,

$$\frac{d\omega_{\pi-\gamma}(E_{\pi},E_{\gamma})}{dE_{\gamma}}
=\frac{2}{\beta_{\pi}E_{\pi}}H[E_{\gamma};\frac{1}{2}E_{\pi}(1-\beta_{\pi}),
\frac{1}{2}E_{\pi}(1+\beta_{\pi})], \eqno(2.15)$$where $\beta_{\pi}\sim 1$, and $H(x;a,b)=1$ ( $a\leq x\leq b$) or $H(x;a,b)=0$
(otherwise), we can analytically calculate the GC-spectrum.
Since $E_p=E_{\pi}^2\exp(-2(a-b))$ and $N_{\pi}=E_{\pi}\exp b$, we have

$$\Phi_{\gamma}(E_{\gamma})=C_{\gamma}\left(\frac{E_{\gamma}}{E_{\pi}^{GC}}\right)^{-\beta_{\gamma}}
\int_{E_{min}}^{\infty}dE_{\pi}
\left(\frac{E_p}{E_p^{GC}}\right)^{-\beta_p}N_{\pi}\frac{2}{\beta_{\pi}E_{\pi}}$$
$$\simeq\left\{
\begin{array}{ll}
2e^bC_{\gamma}(E_{\pi}^{GC})^{2\beta_p}\left(\frac{E_{\gamma}}{E_{\pi}^{GC}}\right)^{-\beta_{\gamma}}\int_{E_{\pi}^{GC}}^{\infty}
dE_{\pi}E_{\pi}^{-2\beta_p}, {\rm ~~~if~}E_{\gamma}\leq E_{\pi}^{GC},\\\\
2e^bC_{\gamma}(E_{\pi}^{GC})^{2\beta_p}\left(\frac{E_{\gamma}}{E_{\pi}^{GC}}\right)^{-\beta_{\gamma}}\int_{E_{\gamma}}^{\infty}
dE_{\pi}E_{\pi}^{-2\beta_p}, {\rm ~~~if~}E_{\pi}^{GC}<E_{\gamma}<E_{\pi}^{max},\\\\
2e^bC_{\gamma}(E_{\pi}^{GC})^{2\beta_p}\left(\frac{E_{\gamma}}{E_{\pi}^{GC}}\right)^{-\beta_{\gamma}}
\exp\left(-\frac{E_{\gamma}}{E_{\pi}^{max}}+1\right)\int_{E_{\gamma}}^{\infty}
dE_{\pi}E_{\pi}^{-2\beta_p}. {\rm ~~~if~}E_{\gamma}\geq E_{\pi}^{max}.\\\\
\end{array} \right. \eqno(2.16)$$Note that the integration intervals are shown in Fig. 4 and
a phenomenological exponential cut factor in Eq. (2.16) describes
the suppression of the energy spectrum at $E_{\gamma}>E_{\pi}^{max}$ due to the gluons at $x<x_c$
have been condensed at $x_c$.

    The integrating results are (see Fig. 5)

$$E_{\gamma}^2\Phi^{GC}_{\gamma}(E_{\gamma})\simeq\left\{
\begin{array}{ll}
\frac{2e^bC_{\gamma}}{2\beta_p-1}(E_{\pi}^{GC})^3\left(\frac{E_{\gamma}}{E_{\pi}^{GC}}\right)^{-\beta_{\gamma}+2} \\ {\rm ~~~~~~~~~~~~~~~~~~~~~~~~if~}E_{\gamma}\leq E_{\pi}^{GC},\\\\
\frac{2e^bC_{\gamma}}{2\beta_p-1}(E_{\pi}^{GC})^3\left(\frac{E_{\gamma}}{E_{\pi}^{GC}}\right)^{-\beta_{\gamma}-2\beta_p+3}
\\ {\rm~~~~~~~~~~~~~~~~~~~~~~~~ if~} E_{\pi}^{GC}<E_{\gamma}<E_{\pi}^{max},\\\\
\frac{2e^bC_{\gamma}}{2\beta_p-1}(E_{\pi}^{GC})^3\left(\frac{E_{\gamma}}{E_{\pi}^{GC}}\right)^{-\beta_{\gamma}-2\beta_p+3}
\exp\left(-\frac{E_{\gamma}}{E_{\pi}^{max}}+1\right).
\\ {\rm~~~~~~~~~~~~~~~~~~~~~~~~ if~} E_{\gamma}\geq E_{\pi}^{max},
\end{array} \right. \eqno(2.17)$$where $E_\pi^{GC}$ is the break energy at the GC-threshold determined by the critical momentum $(x_c,k_c)$. We call Eq. (2.17) as the GC-spectrum

     We establish a relation between $E_{\pi}^{GC}$ and $E_{\pi}^{max}$ in Eq. (2.17). According to Eq. (2.2) and Fig. 2,
the GC plays a role if the GC peak is located at $y_{max}=\ln(\sqrt{S}/k_c)$, i.e.,

    $$x_c=\frac{k_c}{\sqrt{S_{GC}}}e^{-y_{max}}=\frac{k_c^2}{S_{GC}}. \eqno(2.18)$$
 Using $\sqrt{S}=\sqrt{2m_pE_p}$ and Eqs. (2.10) and (2.14), which gives a relation between $E_{\pi}$ and $E_p$, one can get

$$E_{\pi}^{GC}=exp\left(0.5\ln \frac{k^2_c}{2m_px_c}+a-b\right). \eqno(2.19)$$On the other hand, $S_{max}$ in Fig. 2 corresponds to the position where the GC peak disappears in the center of the rapidity $y$, i.e.,

$$x_c=\frac{k_c}{\sqrt{S_{max}}}e^0=\frac{k_c}{\sqrt{S_{max}}}.\eqno(2.20)$$therefore,

$$E_p^{max}=\frac{k_c^2}{2m_px_c^2}. \eqno(2.21)$$ Recycling  Eqs. (2.10) and (2.14) we obtain

$$E_{\pi}^{max}=exp\left(0.5\ln \frac{k^2_c}{2m_px_c^2}+a-b\right). \eqno(2.22)$$Comparing it with Eq. (2.19) we have

$$E_\pi^{max}=e^{b-a}\sqrt{\frac{2m_p}{k_c^2}}\left(E_\pi^{GC}\right)^2, \eqno(2.23)$$where all energies take the GeV unit.

    It is important to note that if the actual accelerated proton energy fails to reach $E_p^{max}$, premature suppression
of the $\gamma$-ray spectra will occur. In such instances, a free parameter $E_{\pi}^{cut}<E_{\pi}^{max}$ will substitute $E_{\pi}^{max}$ in the mentioned above formulas.

    Finally, using Eqs. (2.10) and (2.14) we have a useful relation

$$E_p+m_p=\frac{2m_p}{m^2_{\pi}}E_{\pi}^2.\eqno(2.24)$$

\begin{figure}[htbp]
	\begin{center}
		\includegraphics[width=0.8\textwidth]{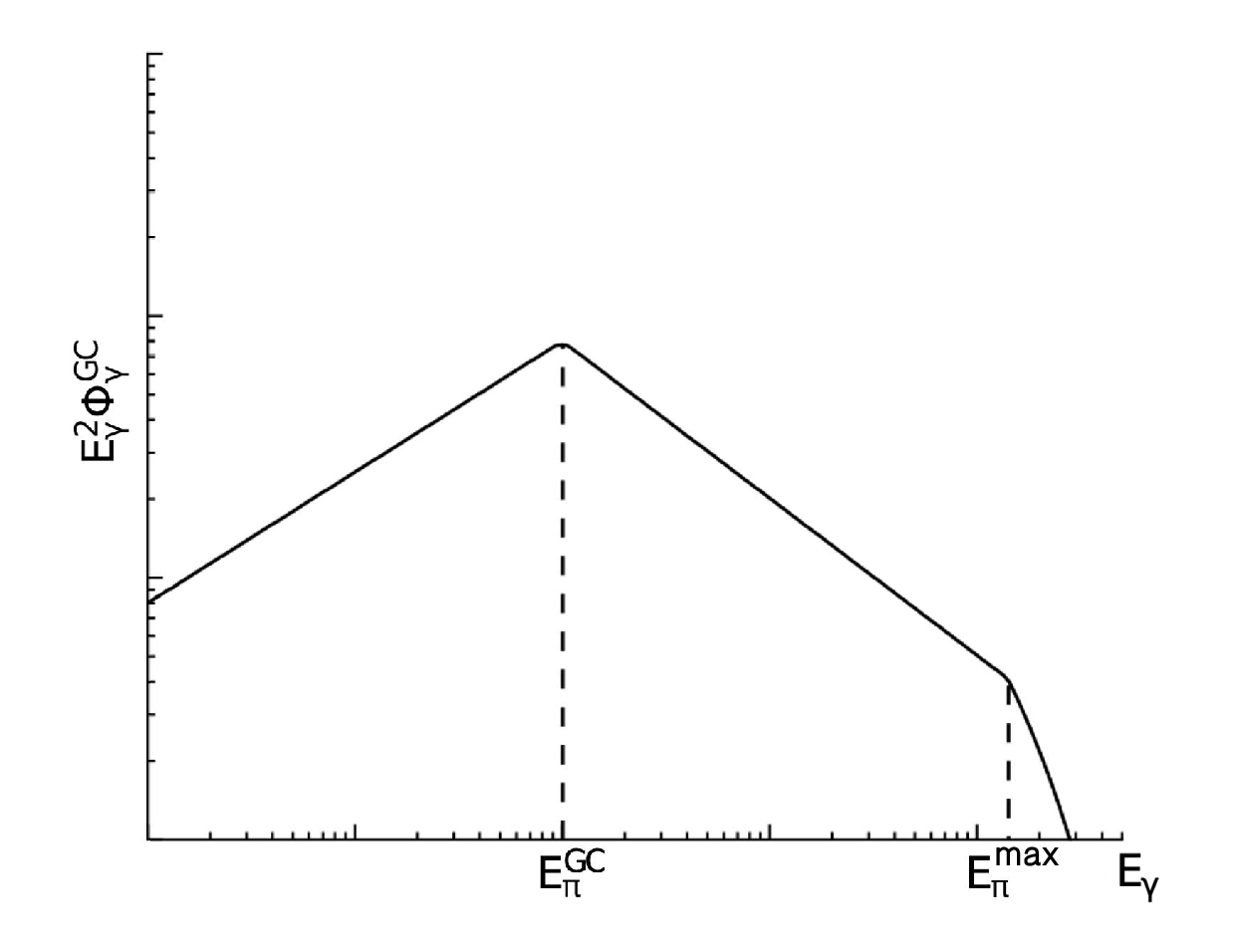} %
		\caption{A schematic GC-spectrum predicted by the hadronic model with the GC-effect.
		}\label{fig:5}
	\end{center}
\end{figure}

   We emphasize once again that Eqs. (2.17), (2.23) and (2.24) are not the commonly used qualitative empirical parameter
formulas in the literature, but rather the analytical solutions of the GC-model. It demonstrates the inherent relationship among different parameters, and these parameters have clear physical meanings in the GC-spectrum shown in Fig. 5.
The GC-spectrum has the following characteristics. (i) The PL at $E_{\gamma}<E_{\pi}^{GC}$ origins
from a steep increase in the $\pi$-number (see Fig. 4), thus the lower limit of integration in Eq. (2.1) is independent of  $E_{\gamma}$, while the PL at $E_{\gamma}>E_{\pi}^{GC}$ is a result of Eq. (2.14); (ii) The gamma signal disappears at $E_{\gamma}>E_{\pi}^{max}$ if the protons are accelerated above $E_p^{max}$. We will show these two points in the following examples.

\section{\bf THE BROKEN POWER-LAW IN GRB 221009A}

     We employ the GC-model to investigate the GRB 221009A event. When analyzing events of this nature, corrections for
the influence of extragalactic background light (EBL) becomes imperative, particularly as high-energy photons traverse vast interstellar distances. Notably, LHAASO has recently provided new intrinsic spectral energy distributions that have undergone these corrections, covering the time span from 230 to 900 seconds after the trigger [25]. Regrettably, the available LHAASO data are constrained to energies above 200 GeV. However, gamma-ray spectra at lower energies for GRB 221009A have been captured by Fermi-LAT over the initial 2000 seconds following the Fermi-GBM trigger [26]. It is worth noting that data within the first 300 seconds from Fermi-LAT may potentially saturate the detectors [27]. Consequently, we exclusively utilize the LHAASO data during the interval of 300-900 seconds to seamlessly connect with the Fermi-LAT data, as illustrated in Fig. 6.

   \begin{figure}[htbp]
   	\begin{center}
   		\includegraphics[width=0.8\textwidth]{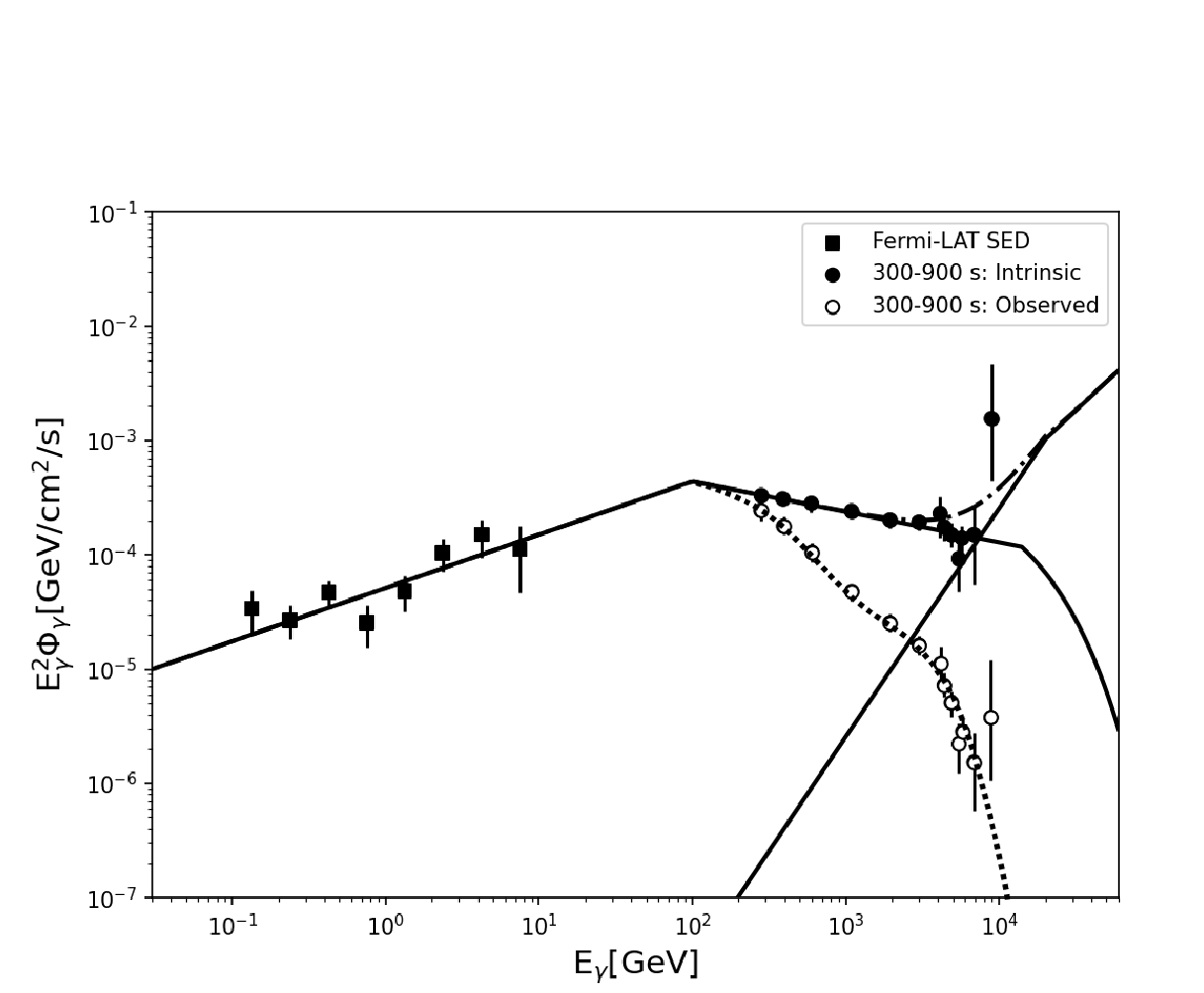} 
   		\caption{ An intrinsic spectral energy distribution of GRB 221009A compares with the GC-spectrum for the p-heavy nuclei, pp collisions (the solid curve) and sum of the of them (point-broken curve), where the contributions of
   the EBL attenuation are deducted from the observed spectra (the dashed curve) using the model of Saldana-Lopez et al [28].
            Data are taken from LHAASO in [300-900]s
   			[25] and Fermi-LAT in [300,900]s [26]. The parameters: for the p-heavy nuclei collisions,
   $C_{\gamma}=3.09\times 10^{-12}(GeV^{-2}cm^{-2}s^{-1})$, $\beta_{\gamma}=1.53$, $\beta_p=0.867$, $E_{\pi}^{GC}=100~GeV$ and $E_{\pi}^{cut}=14~TeV$; for the $pp$-collisions,  $C_{\gamma}=9\times 10^{-19}(GeV^{-2}cm^{-2}s^{-1})$, $\beta_{\gamma}=0$, $\beta_p=0.867$ and $E_{\pi}^{GC}=20~TeV$. We noticed that the work [25] used the traditional hadronic model to fit this spectrum.
   Due to the fact that the GC model describes $N_{\pi}$ using far fewer free parameters than that of the traditional hadronic model,
   the former should be have better fitting goodness.
  }\label{fig:6}
   	\end{center}
   \end{figure}

     Figure 6 presents the spectral energy distribution of GRB 221009A and its comparison with the GC-spectrum. The value
of $E_\pi^{GC}$ relies on the baryon number $A$ of the target nucleus in the $pA$ (or $AA$) collisions. For instance, $E_\pi^{GC}$ is ${\sim}20~TeV$ in the $pp$-collisions, and it is $100~GeV$ for the $p$-heavy nucleus collisions [17].  Note that the broken point at $\sim 100~GeV$ in Fig. 6 is also consistent with the previous fitting GRB 180720B and GRB 190114C in [20].

   The observational precision of LHAASO has seen significant improvement owing to the extensive signals collected by
its expansive detector array. The notable alignment between the linear segment of the GC-spectrum and the LHAASO data is particularly striking, even accounting for the exceptionally small statistical errors. The PL behavior of the meson yield in Eqs. (2.10) and (2.14) outcome serves to reinforce the credibility of the saturation assumption in $N_{\pi}$.

     The LHAASO data reveal that gamma-ray events beyond $10$ TeV in the intrinsic spectrum of GRB221009A do not exhibit
softening. This phenomenon is also observed in the VHE gamma-ray spectra of some AGNs [20].
The observation of the hardening phenomenon in TeV gamma-rays from cosmological distances carries significant implications for potential new physics beyond the standard model. For instance, it has been interpreted as originating from axions [29] or Lorentz invariance violation [30]. Axions are hypothetical particles introduced by theoretical models beyond the Standard Model (SM). These unobserved particles couple with two photons, leading to a suppression of EBL absorption of high-energy gamma-rays and increased transparency of space. Alternatively, if the threshold for interacting with EBL is slightly increased due to Lorentz invariance violation, the process $\gamma\gamma\rightarrow e^+e^-$ is suppressed, resulting in a more transparent universe. These scenarios highlight the potential for TeV gamma-ray observations to provide insights into novel physics beyond our current understanding.

    The GC-model may provide a realistic explanation for the observed hardening VHE gamma-ray spectra.
In Fig. 6, we present a GC-spectrum represented by the dashed lines, derived from the $pp$ collisions, with $\beta_{\gamma}\sim 0$ indicating negligible energy loss of gamma-rays in the $pp$-collision environment [20].

    The hardening gamma spectrum caused by GC can occur in different types of processes, as long as there are
high-energy $pp$ collisions there. In fact, similar spectra have been observed in the energy spectra of various celestial objects, including 1ES 0229+200, 1ES 1101-232, 1ES 0347-121, PKS 2005-489, H1426+482, and MKn 501 (see Fig. 8 in [20]). The GC-spectra near $E_{\pi}^{GC}\sim 20TeV$ imply a necessity for proton acceleration beyond $E_p^{GC}\sim 40~EeV$. It is noteworthy that a recent detection of an extremely energetic cosmic ray at 244~EeV by the surface detector array of the Telescope Array experiment [31] is of particular relevance. Taking into account the Greisen-Zatsepin-Kuzmin (GZK) suppression, this observation suggests that EeVatrons in extragalactic sources could potentially serve as the sources of the GC-spectrum beyond the TeV energies, we will discuss it in detail elsewhere.

    We make a comparison of the radiation powers between the hadronic scenario without the GC-effect and with the
GC-effect. In the former case, a significant portion of the kinetic energy in hadronic collisions typically transforms into heat for secondary particles in the central region of the collisions. Consequently, the meson yield follows a logarithmic dependence, i.e., $N_{\pi} \propto \ln (\sqrt{S}/m_{\pi})$, where $\sqrt{S}$ is the interaction energy in the center-of-mass (C.M.) system.
On the other hand, the GC-effect maximizes the utilization of available kinetic energy to generate new particles, resulting in a meson yield proportional to $\sqrt{S}/m_{\pi}$. Consequently, the ratio of $N_{\pi}$ with the GC-effect to $N_{\pi}$ without the GC-effect is much larger than 1. This outcome underscores that the GC-effect facilitates the efficient conversion of collision kinetic energy into radiation energy. This is one of the reasons why GRB displays strong radiation power.

\section{THE COLOR PENETRATION IN GRB 170817A}

    The distribution of condensed gluons within a nucleon not only reveals a pronounced peak at $(x_c, k_c)$
but also indicates the absence of gluons at $x < x_c$. Let's consider a neutron (n) incident on a nucleus (A). The average penetration distance ($L$) of the neutron is defined as $L = 1/(\sigma N_A)$, where $N_A$ represents the nuclear density. The cross section $\sigma(\sqrt{S})$ encompasses contributions from both elastic and inelastic scattering, as well as resonance excitation.
In the parton picture [33-35], at high energy, valence quarks tend to traverse with a loss of only approximately $2\%$ of their initial momentum. These valence quarks give rise to outgoing hadrons in the fragmentation region. Interactions between small partons from the two incident hadrons in the rapidity central range dominate the cross section. The resistance between two colliding nucleons is contingent upon the multiplicity and strength of interactions among their partons. Consequently, the predominant contribution to the cross section $\sigma$ arises from the exchange of small partons between the two colliding nucleons, particularly since these small partons, mainly gluons, exhibit large multiplicity  and low momentum transfer.

   Specifically, the cross section $\sigma(pp\rightarrow X)$ at high energy is dominated by
the production of gluon mini-jet $\sigma(\sqrt{S})\propto N_g(\sqrt{S})$. Note that $\sigma(\sqrt{S}>\sqrt{S_{max}})\rightarrow 0$ due to Fig. 4. It means that $L\rightarrow \infty$ at $\sqrt{S}>\sqrt{S_{max}}$ and the nucleus becomes traversable.
Therefor, this pair of neutron stars can maintain spherical merging with almost no resistance and an explosion occurs at the maximum spherical overlap. We call it the color penetration. $\sqrt{S_{max}}$ is its threshold, which relates to the value of $x_c$.
For example, for a heavy nucleus, its $E_{\pi}^{GC}\sim 100~GeV$ and $\sqrt{S_{max}}\sim 1~PeV$. Thus, the color penetration
exists in the extra high energy hadronic collisions and it is an extremely rare event.
In reference [15] it was predicted that this effect may impede the increase of new particle events in future ultra high energy hadron colliders.

     However, the binary neutron star merger scenario may change this situation. The core $A^*$ of a neutron star is
a huge nucleus composed of nearly an infinite number of neutrons, making the GC-critical value $x_c$ significant increases
in such a large nucleus [17]. In principle, $x_c$ approaches 1 as $A$ approaches infinity. However, the kinematical restriction leads to a maximum $x_c<1$ at $A\rightarrow \infty$. Let us detail it.
Since pions dominate the secondary particles in $pp\rightarrow X$, we consider $\sigma(pp\rightarrow \pi)$.
Energy conservation demands $E_p\geq m_p$ and $E_{\pi}\geq m_{\pi}$ in Eq. (2.11). Therefore,
$E_{\pi,min}^{GC}=0.135~GeV,~~E_{p,min}^{GC}=0.938~GeV$, which corresponds to $E_{\pi}^{max} = 0.255~GeV$,
or $\sqrt{S_{max}}=3.7~GeV$ and $x_c=0.27$ if $k_c\sim 1~GeV$.
It implies that two neutrons from two star cores at the C.M. system have energies $E_{c.m.}+E_{c.m.}=3.7~GeV$ and the
Lorentz factor $\gamma=1.85/0.938=2$ or $\beta=v/c=0.87$.

    If we consider the neutron star in GRB 170817A with a radius of approximately $R \sim 20~km$ and a spin
frequency of $f \sim 2000/s$, each neutron during the merging process exhibits a $\beta \sim 0.67$. Taking into account the additional contribution of the star orbit rotation, the phenomenon of color penetration becomes plausible. When a collision manages to breach the shell barrier of neutron stars and penetrate the core area from both sides, a notable reduction in the cross-sectional area $\sigma$ is observed. This suggests that binary neutron stars can merge with minimal resistance until two neutrons reach each other's boundaries. This scenario has the potential to result in a spherically symmetric explosion source.
However, it's important to note that not all binary neutron star mergers exhibit the color penetration effect unless they possess a sufficiently high collision energy.

    The color penetration observed in GRB 170817A is accompanied by a characteristic gamma-ray spectrum.
Cosmic gamma-rays typically arise from the $pp\rightarrow \pi^0\rightarrow 2\gamma$ process within the hadronic scenario. This process is highly affected by the gluon distributions in nucleons. The GC-effect may give rise to the typical BPL with an exponential suppression factor in gamma-ray energy spectra. Besides, in a nucleon-nucleon collision,
$\Phi^{GC}_{\gamma,max}(pA)/
\Phi^{GC}_{\gamma,max}(A^*A^*)\sim \sqrt{S_{max}(pA)}/\sqrt{S_{max}(A^*A^*)}\gg 1,$ since $E_{\pi}^{max}(pA)\gg
E_{\pi}^{max}(A^*A^*)$. Therefore, the GC-peak at the MeV band in the merger of binary neutron stars is significantly reduced.

    It is noteworthy that no hard gamma-rays were detected by Fermi-GBM [36] and INTEGRAL [37] during the merger time of
GW 170817. Despite subsequent analyses of data recorded by high-energy ($> 100MeV$) and VHE ($> 100GeV$) gamma-ray detectors, such as Fermi-LAT, no instances of high-energy counterparts conclusively associated with the event were identified [38,39].
An interruption in data collection by Fermi-LAT during the trigger moment of GW 170817, caused by its passage through the South Atlantic Anomaly, prevented an assessment of the presence or absence of high-energy emission. Nevertheless, negative observations were recorded with H.E.S.S., a detector that commenced observations merely five hours after GW 170817. The GC-model predicts that the gamma-ray spectrum of GRB 170817A exhibits an isolated, narrow, and weak spectrum in the MeV range.

    An energy range approximately centered around $\sim 200~MeV$ represents an intermediate zone, positioned between the KeV
and GeV ranges, which has received limited attention in investigations of GRB emissions. This is attributed to both the characteristics of scintillation detectors used for GRB observations, where the efficiency of photon detection in this range is insufficient, and the steeper and narrower BPL shape of the GC-spectrum. Fortunately, a new space-based gamma-ray telescope, GAMMA-400, has been designed to continuously measure within the 20 MeV to several TeV energy range [40]. The exceptional characteristics of this instrument make it a potent tool for identifying GC-spectral signatures at MeV in short GRB events. Obviously, our discussion on the GRB 170817A event is only a preliminary qualitative analysis. Further research using
fluid dynamics is necessary.

\section{DISCUSSIONS AND CONCLUSION}

    (i) Most GRBs are detected below a few MeV. However, the GC-spectra Eq. (2.17) may span a broad VHE range and precise
measurements are required to distinguish the GC-model from other theoretical models. A new electromagnetic window in the VHE domain
was just opened recently by ground-based imaging atmospheric Cherenkov telescopes.  Unfortunately, GRBs are rare and highly directional events in the universe, occurring over extremely short time periods. It is a low-probability event to observe such cosmic ray events simultaneously and completely with high precision. GRB221009A is currently the best-recorded event we have collected. The GRB 221009A spectrum of such a complete GeV-TeV energy band is exceptionally rare, with an estimated probability of occurrence at one in
ten thousand years [41], and of course, it is a rare opportunity to test the GC-model. We have previously discussed GRB 190114C and GRB 180720B in [20], but it is evidently difficult to make conclusions due to the scarcity of observational data. We look forward to encountering GRB events similar to GRB 221009A in the future.

    Of course, a few examples alone are not enough to judge a model. Fortunately, GC occurs at the most fundamental level
of the proton. The GC-spectra should also be widely displayed in other high-energy gamma spectra. In fact, the observed BPL form in gamma-ray spectra is not exclusive to GRB 221009A, it has also been documented in active galactic nuclei (AGNs) [20],  pulsars [42,43] and Galactic center [44]. In fact, more than fifty such events have been cited by the GC-model. These examples imply that the hadronic scenario involving the GC-effect is a prevalent phenomenon in nature, albeit with varied manifestations in different environments. For instance, in a GRB event, gamma-ray emissions occur when particle jets collide with the remnants of precursor stars, leading to prompt emissions over a short duration. In contrast, in AGNs, similar collisions may persist for an extended period due to the continuous production of a large number of PeV protons through potent acceleration mechanisms, although our understanding of these mechanisms remains incomplete. This diversity in manifestations underscores the complexity and richness of the hadronic scenario associated with the GC-effect across different astrophysical contexts.

     (ii) Comparing GC with Bose-Einstein condensation (BEC), we observe that, despite both involving multi-bosons
sharing a single wave function, they represent distinct physical phenomena. Thus, GC opens a new window into nature. For example, the GC-model efficiently generates electromagnetic radiation in the universe. While positive and negative electron annihilation undoubtedly maximizes photon production efficiency, such events are rare in nature. Inverse Compton scattering is a common mechanism for explaining VHE gamma-ray spectra, but it requires an ample supply of soft photons to serve as targets for scattering electrons.
Furthermore, GC represents a first instance of the butterfly effect occurring in elementary particles. Consequently, concerns about the efficacy of perturbation treatment are unnecessary, as once the CGC exhibits a strong chaotic phenomenon, it proves challenging to suppress. Additionally, the $\pi$-cluster, as an intermediate state in the GC-model, closely resembles $\pi$-condensation, a phenomenon anticipated in nuclear physics. This warrants further investigation.

    (iii) The maximum energies in $p-Pb$ and $Pb-Pb$ collisions at the LHC are 8.16 TeV and 5.02 TeV,
respectively. Additionally, the Auger collaboration indirectly used cosmic ray data at the top of the atmosphere and found that the $pA$ cross section at $\sqrt{s}\sim 100TeV$ is a normal value of $\sim 567mb$ with no significant increment. An important question arises: why have they not recorded the GC-effect? We consider $E_{\pi}^{GC}(Pb-Pb)=100GeV$ and $E_{\pi}^{GC}(p-light~nuclei)=1TeV$ according to Fig. 11 in [17]. The results are $\sqrt{S_{Pb-Pb}^{GC}}=1TeV$, $\sqrt{S_{Pb-Pb}^{cut}}=100TeV$, $\sqrt{S_{p-light~nuclei}^{GC}}=10TeV$, and $\sqrt{S_{p-light~nuclei}^{cut}}=10~PeV$. Since these estimations are based on central collisions, the actual collision threshold should be larger than these estimations.
Therefore, we caution that further increases in hadron collision energies in the upcoming LHC plans may lead to unexpectedly intense gamma-rays in the accelerator, resembling artificial mini-GRBs, and potentially causing damage to the detectors [17]. Moreover, GC raises the possibility of a sudden disappearance of gluon distributions, potentially halting the increase of new particle events in an ultra-high-energy heavy-ion collider [15].

    (iv) We have noticed some attempts to study GRB beyond the traditional hadronic scenario.  For example,
work [45] uses the magnetohydrodynamic simulation tracking the magnetosphere of a collapsing magnetar. The results suggest a novel GRB scenario, which creates a delayed high-energy counterpart of the merger gravitational waves;
work [46] presents a mechanism based on internal self-annihilation of dark matter accreted from the galactic halo in the inner regions of neutron stars that may trigger full or partial conversion into a quark star. This mechanism differs in many aspects from the most discussed scenario associating short GRBs with compact object binary mergers; Work [47] simulates neutron-star-to-quark-star burning at stellar scales and estimate the resulting energy release and ejecta, which can help us to understand GRBs. Although the GC-model alon is also a new attempt, but it is a natural development of the mature traditional hadronic scenario, where
the corrections of the GC-effect to the cross section of $pp$-collisions are considered. The GC-model employs fewer free parameters and assumptions, while describing the spectral energy distribution that encompasses the richest information from the radiation source. Additionally, based on the parameters obtained from fitting the GC-model to various cosmic ray spectra, we anticipate the direct observation of the GC-effect in the next-generation high-energy hadron collider, without the need for special detectors. Therefore, the
GC-model is a topic worth our attention.

    To summarize, we illustrate a simple logical relationship between GC in the proton and the GRB spectra during
the burst phase. When the CGC gluons undergo evolution through the singularity of a nonlinear QCD evolution equation, these gluons will enter a chaotic state. The resulting intense oscillations in the gluon distribution may lead to strong antishadowing, causing a large number of soft gluons to condense at a critical momentum. On the other hand, during the burst phase of a GRB event, protons are accelerated above the GC-threshold, leading to a sudden influx of condensed gluons into the proton-nuclei collisions, giving rise to the observed bump in the gamma-ray spectrum. Moreover, as long as these pion yield saturates, energy conservation and relativistic covariance directly result in the BPL form the gamma spectrum, as observed by Fermi-LAST and LHAASO. Besides, the remarkably symmetrical explosion cloud in kilonova AT2017gfo and the absence of a very high-energy gamma-ray signature in GRB 170817A also can qualitatively explained by the above GC-model.

\paragraph{\bf Acknowledgments}  We thank Zi-Qing Xia for the useful comments. We also thank Qi-hui Chen and Wei Kou for the help in the calculations.

\end{document}